\documentstyle[11pt,iau185,twoside,epsf]{article}

\begin{document}
\title{Djehuty: A Code for Modeling Whole Stars in Three Dimensions\altaffilmark{1}}
\author{S. Turcotte\altaffilmark{2}, G. Bazan, J. Castor, R. Cavallo,
H. Cohl, K. Cook, D. S. P. Dearborn, D. Dossa, R. Eastman, P. P. Eggleton, P. Eltgroth, S.
Keller, S. Murray, A. Taylor}
\affil{Lawrence Livermore National Laboratory, Livermore, CA, USA}
\altaffiltext{1}{This work was performed under the auspices of the U.S.
Department of Energy, National Nuclear Security Administration by the University of
California, Lawrence Livermore National Laboratory under contract No. W-7405-Eng-48.}
\altaffiltext{2}{e-mail: sturcotte@igpp.ucllnl.org}

\begin{abstract}
The DJEHUTY project is an intensive effort at the Lawrence Livermore
National Laboratory (LLNL) to produce a general purpose
3-D stellar structure and evolution code to study dynamic
processes in whole stars.
\end{abstract}
 
\section{Introduction}
Stellar models in 1-D work remarkably well for most stars.
However, stars are three dimensional objects and the computing power is now at a
point where we can do better than 1-D models for modeling the large
array of physical processes occurring in stars 
for which spherical symmetry is no longer a valid
approximation. With a 3-D stellar code one can tackle the problems
linked to rotation, turbulent motions and convection, magnetism, binarity,
and explosive phases of stellar evolution in a consistent and
physically meaningful way.

The DJEHUTY code is an evolution of a
radiation hydrodynamics code developed over decades at LLNL.
It is our goal to provide the astrophysical community with the first
general purpose 3-D stellar structure and evolution code 
suitable to study the whole gamut of dynamical processes occurring in
stars.

\section{The DJEHUTY code}
At the heart of DJEHUTY is an Arbitrary Lagrangian-Eulerian (ALE; Barton 1985) 
code for radiation hydrodynamics which treats radiation transport in the
diffusion approximation.
Microscopic physics appropriate for stars has been added. The opacities
used are those of OPAL (Iglesias \& Rogers 1996) for high temperatures 
and Alexander (Alexander \& Ferguson 1994) for low
temperatures. A set of Planck opacities computed from the OP data
(Seaton et al. 1994)
is being used within the hydro code to couple the radiation to the
matter. An improved version of the EFF equation of state (Eggleton et al
1973) is implemented. 
Full nonaxisymmetric gravity will be implemented shortly 
using a Heine boundary solver combined with 
Djehuty radiation diffusion
solvers being used for solving Poisson's equation.
Finally, a general nuclear network (Dearborn~1992) is operational. 
Eventually, a chemical network may be added and a more general
calculation for opacities should be implemented. 

The star is divided in seven logical cubes. Six of the cubes are arranged
and deformed into a spherical (or spheroidal in the general case) shell 
around a central cube. The central cube is subdivided in $N^3$ elements
and the outer cubes are subdivided in $N^2\times$ an arbitrary 
number of radial elements, typically $2N$
in calculations done so far. This mesh structure was chosen to avoid
singularities and poles. 
As the numerical scheme is currently explicit,
time steps are limited by the Courant condition but an implicit method
is being investigated.

The starting model for computations is a 1-D model mapped on the 3-D
grid. Since the 3-D mapping does not result in a perfect hydrostatic
3-D star, the model readjusts itself. Tests show that a new equilibrium
is attained quickly.
As the star evolves, convection and other hydrodynamical processes will
establish themselves naturally.

A complete paper detailing the physics and numerics of the code is in
preparation (Dearborn et al.~2002).

\section{First results and Future Science}

Apart from test problems, the code has so far been applied to a
4~M$_\odot$ star on the main sequence in the goal of studying
core convection and overshoot. This issue is well suited to the code
as it stands now as our maximum spatial resolution is achieved in the
stellar core. It is also one of the more fundamental theoretical
problems in modern astrophysics. Preliminary results show the start of
convective motions which do extend outside the limit of the convective
cores that are typical of 1-D models.
Some models with rotation have been computed but the results are too
preliminary so far to draw any conclusion on the interaction of convection and
rotation. 

Close binaries will be the subject of our efforts in the near future. 
Current plans call for a point-like secondary as a first effort followed
by calculations where both members of the system are in the computational
domain, including dynamics in both stars as well as mass transfer.

\end{document}